\documentclass{article}
\usepackage{spconf,amsmath,graphicx}
 \usepackage{mathtools}
 \usepackage{graphicx}
\usepackage{comment}
\usepackage{xcolor}
\usepackage{graphicx}
\usepackage{subcaption}
\usepackage{booktabs}
\usepackage{enumitem}

\usepackage{multirow}
\usepackage{hyperref}
\usepackage{amssymb}
\usepackage{footmisc}

\newcommand\redsout{\bgroup\markoverwith{\textcolor{red}{\rule[0.5ex]{2pt}{0.4pt}}}\ULon}

\title{Improving fairness in speaker verification via \mbox{Group-adapted Fusion Network}}

\name{\parbox{\textwidth}{\centering
Hua Shen$^{1,2,}$\sthanks{Equal contribution.}, Yuguang Yang${^{2,*}}$, Guoli Sun${^{2}}$, Ryan Langman${^{2}}$, Eunjung Han${^{2}}$, \\
Jasha Droppo${^{2}}$, Andreas Stolcke${^{2}}$}}

\address{$^1$The Pennsylvania State University, $^2$Amazon Alexa AI \\
$^1$huashen218@psu.edu, $^2$\{yuguay, guols, rlangman, cehan, drojasha, stolcke\}@amazon.com
}
\begin{document}
\ninept
\sloppy

\maketitle
\begin{abstract}
Modern speaker verification models use deep neural networks to encode utterance audio into discriminative embedding vectors. During the training process, these networks are typically optimized to differentiate arbitrary speakers. This learning process biases the learning of fine voice characteristics towards dominant demographic groups, which can lead to an unfair performance disparity across different groups. This is observed especially with underrepresented demographic groups sharing similar voice characteristics. In this work, we investigate the fairness of speaker verification models on controlled datasets with imbalanced gender distributions, providing direct evidence that model performance suffers for underrepresented groups. To mitigate this disparity we propose the group-adapted fusion network (GFN) architecture, a modular architecture based on group embedding adaptation and score fusion. We show that our method alleviates model unfairness by improving speaker verification both overall and for individual groups. 
Given imbalanced group representation in training, our proposed method achieves overall equal error rate (EER) reduction of 9.6\% to 29.0\% relative, reduces minority group EER by 13.7\% to 18.6\%, and results in 20.0\% to 25.4\% less EER disparity, compared to baselines.
The approach is applicable to other types of training data skew in speaker recognition systems.

\end{abstract}

\begin{keywords}
speaker verification, model fairness, embedding adaptation, score fusion.
\end{keywords}

\section{Introduction}
\label{sec:intro}

A speaker verification system answers the question of \textit{who is speaking} based on a recording of a spoken utterance. 
With smart home and mobile applications becoming more ubiquitous, speaker verification systems are playing an important role in enabling convenient and secure access to personalized services through natural conversational interactions, such as playing one's favorite music, checking one's calendar, and conducting financial transactions via voice commands.
Users need to be able to count on such personalization features working reliably regardless of the speaker's linguistic or demographic background.

Modern deep speaker verification models, such as d-vectors \cite{heigold2016end, wan2018generalized,nagrani2017voxceleb, chung2018voxceleb2}, x-vectors \cite{snyder2017deep, snyder2018x}, and other variants \cite{li2017deep, garcia2020magneto, desplanques2020ecapa},  are typically trained on large datasets to minimize average speaker identification loss. Such a training paradigm can cause models to overlook distinctive voice characteristics for underrepresented groups (such as gender groups, nonnative speakers, or regional accents) in the training data.
The resulting  lower verification performance for minority groups affects their fair access to services enabled by voice verification technologies. Meanwhile, common speaker verification performance metrics typically measure the overall model performance across all speakers and do not reflect equity of performance over different demographic groups. Prior work has reported that insufficient training data from minority demographic groups could impair performance fairness in state-of-the-art automated speech recognition systems and speaker verification models~\cite{koenecke2020racial, fenu2021fair}.
Similar fairness issues have also been identified in other areas~\cite{mehrabi2021survey}, such as face recognition~\cite{wang2019racial} and recommender systems \cite{beutel2019fairness}.

A common way to improve model fairness is to collect more annotated training data from minority groups, which can be prohibitively expensive and time-consuming.
Here we propose an algorithmic approach to overcome fairness issues arising from typical speaker verification systems, consisting of two major components, jointly called group-adapted fusion network (GFN). First, we use group-wise embedding adaptation to improve the front-end embedding encoder's ability to extract better discriminative features within a demographic group with similar voice characteristics. Second, we fuse scores from different embedding encoders via learnable weights to improve generalization and prevent overfitting. Embedding adaptation has been applied in few-shot learning or transfer learning settings to refine task-specific features, with applications in computer vision \cite{ye2020few,li2019finding} and speech \cite{Tan2021ImprovingSI}, among others.

We illustrate the fairness problem and demonstrate the effectiveness of our solution for the case of gender imbalance in the training data. {By constructing training sets with various degrees of imbalance, as well as metrics for performance disparity, we aim to systematically probe model unfairness, understand its causes, and offer a generalizable solution.} 
Although our approach is only evaluated on imbalanced gender groups, it is applicable to other demographic groups (\emph{e.g.,} children, the elderly) affected by underrepresentation.

Our main contributions are as follows: We provide direct evidence that imbalanced group representation in speaker recognition training sets can lead to model unfairness. We propose a general, modular architecture based on group embedding adaptation and score fusion that alleviates model unfairness. Our approach also comes with a set of tools to rigorously inspect, evaluate, and analyze  model unfairness and proposed solutions. Finally, the training and evaluation datasets used in this work are available for other uses.\footnote{\scriptsize\url{https://github.com/huashen218/Voxceleb-Fairness.git}\label{data_url}}

\vspace{-1em}
\section{Group-adapted Fusion Network}

\begin{figure*}[t]
    \centering
    \includegraphics[width=0.9\textwidth]{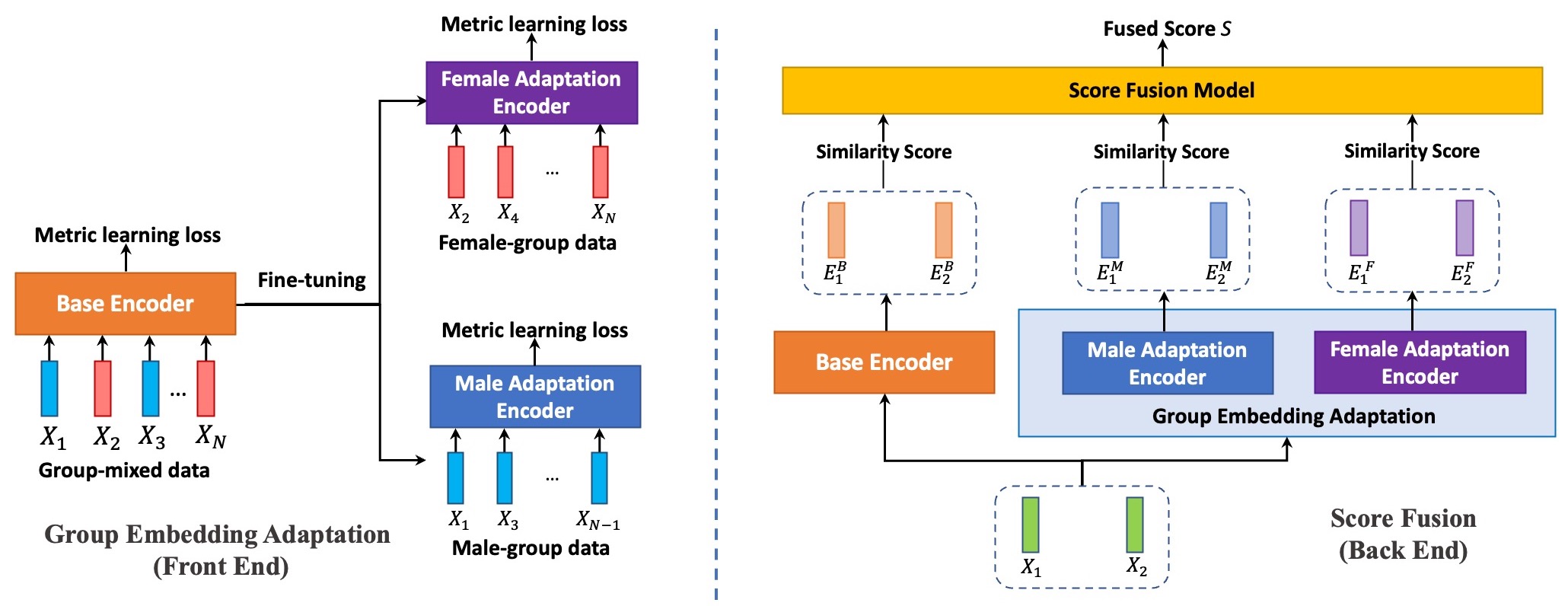}
    \caption{The overview of proposed group-adapted fusion network, which consists of the front-end group adaptation encoders (\textbf{left}) to extract group-wise embeddings and the back-end score fusion model (\textbf{right}) to fuse scores from all embedding encoders.}
    \label{fig:network_diagram}
    \vspace{-1.5em}
\end{figure*}

The architecture of the group-adapted fusion network is shown in Figure~\ref{fig:network_diagram}. It consists of front-end encoders to extract base and group-adapted embeddings and a back-end score fusion model to fuse base scores and group-specific scores to generate the fused score. The core idea is motivated by ensemble learning \cite{zhou2021ensemble} and mixture-of-experts \cite{jordan1994hierarchical,shazeer2017outrageously}, where multiple expert networks model complementary data characteristics and are fused at the score level. Another precedent is speaker verification based on adaptation transforms that are gender-specific but applied uniformly to all speakers~\cite{StolckeEtAl:ieee-salp2007}.

\vspace{-.5em}
\subsection{Group Embedding Adaptation}
\vspace{-.5em}

The base and group-adapted encoders are three separate deep neural networks with the same architecture (ResNet-34 variants \cite{chung2020defence}). All are trained with metric learning objectives~\cite{chung2020defence}. We first train the base encoder with gender-mixed data to capture generic voice characteristics, and then fine-tune the pre-trained base encoder with gender-specific training data. In the training stage, inputs to each encoder are mini-batches (batch size $N$) of audio features $X_1,X_2,...,X_N \in R^{T \times F}$ (e.g., log Mel filter banks), where $T$ is the number of audio frames and $F$ is the feature dimension. Outputs from each encoder are length-normalized embeddings $E\in R^D$, where $D$ is the embedding dimension. Embeddings are fed into the metric loss function for network training. 
We use $E_i^B, E_i^M, E_i^F, i=1,..., N$ to denote base embeddings, \textbf{female-adapted} embeddings and \textbf{male-adapted} embeddings produced from corresponding encoders.

\vspace{-.5em}
\subsection{Score Fusion}
\vspace{-.5em}

We next leverage a score fusion model to aggregate the 
three embeddings, to predict if an utterance pair $({X}_1, {X}_2)$ is from the same speaker.
In the score fusion stage, the inputs are utterance-level base embedding pairs $({E}^B_1, {E}^B_2)$, female- and male-adapted embedding pairs $({E}^F_1, {E}^F_2)$ and $({E}^M_1, {E}^M_2)$.
First, we compute cosine similarities $S^B, S^F, S^M$ between pairs for base, female-adapted and male-adapted embeddings, respectively:

\vspace{-1.2em}
\begin{equation} 
\label{eq:fusion}
\begin{split}
S^B &= \text{CosineSimilarity}({E}^B_1, {E}^B_2),  \\
S^F &= \text{CosineSimilarity}({E}^F_1, {E}^F_2),\\
S^M &= \text{CosineSimilarity}({E}^M_1, {E}^M_2) \\
S &= \text{Sigmoid}(f([S^B, S^F, S^M]; W)).
\end{split}
\end{equation}

We then use a score fusion model $f(\cdot)$ to aggregate the three similarity scores into the fused score for speaker verification.
We employ a multilayer perceptron (MLP) $f(\cdot)$  with the three similarity scores $[S^B, S^F, S^M]$ as inputs and $W$ as the learnable model weights.
To train the score fusion model, we construct positive and negative training pairs from the training set for contrastive learning. 
Positive utterance pairs $\mathcal{P}$ are sampled from the same speaker; negative pairs $\mathcal{N}$ are formed by sampling utterances from different speakers of both same and different gender. We train the fusion model with binary cross-entropy loss
\begin{equation} 
\label{eq:loss}
\begin{split}
L = - \frac{1}{M} \left(\sum_{n\in \mathcal{P}}y_n\log S_n+\sum_{n\in \mathcal{N}} (1-y_n)\log (1-S_n) \right)
\end{split}
\end{equation}
where $S_n$ is the fused score output of the $n$-th utterance pair $({X}_1, {X}_2)_n$, $M= (|\mathcal{P}| + |\mathcal{N}|)$ and $y_n$ is the corresponding label ($y_n=1$ indicates the paired utterances are from the same speaker, $y_n=0$ otherwise).

During inference, given a pair of utterances for verification, we first extract their base, female- and male-adapted embeddings from the three encoders. The three embeddings are fed into the score fusion model producing score $S$. If $S$ is greater than a predefined threshold, we predict that two utterances are from the same speaker; otherwise, they are deemed from different speakers.

\section{Data and Experiments}
\label{sec:data_eval}

\noindent \textbf{Data.} We use VoxCeleb1 \cite{nagrani2017voxceleb} and VoxCeleb2 \cite{chung2018voxceleb2} datasets, which have gender information for each speaker, to construct customized training and evaluation datasets.
We constructed training datasets with different gender ratios from subsets of the  VoxCeleb2 dataset. These subsets have the same total of 2,500 speakers but with different numbers of male and female speakers. The female-to-male (F:M) gender ratio ranges from 9:1 to 1:9, as shown in Table~\ref{table:training_data}. We will refer to them as \textbf{VoxCeleb2-GRC} (gender ratio controlled) datasets.
To accurately evaluate model fairness based on gender, we also constructed an evaluation dataset based on VoxCeleb1. We call it the \textbf{VoxCeleb1-F} (Fairness) dataset. VoxCeleb1-F strictly controls for the presence of positive and negative trials with same or different genders, as shown in Table~\ref{table:eval_data}.  We use F and M to denote the female and male groups, respectively. 

\begin{table}[t]
\caption{VoxCeleb2-GRC datasets with different gender ratios.}
\centering
\scriptsize
\begin{tabular}{@{}c|cc|cc@{}}
\toprule
\begin{tabular}[c]{@{}c@{}}\textbf{Gender} \\ \textbf{ratio F:M}\end{tabular} & \begin{tabular}[c]{@{}c@{}}\textbf{Female}\\ \textbf{speakers}\end{tabular} & \begin{tabular}[c]{@{}c@{}}\textbf{Male}\\ \textbf{speakers}\end{tabular} & \begin{tabular}[c]{@{}c@{}}\textbf{Female}\\ \textbf{utterances}\end{tabular} & \begin{tabular}[c]{@{}c@{}}\textbf{Male}\\ \textbf{utterances}\end{tabular} \\ \midrule
9:1 & 2,250  & 250  & 387,322  & 45,181  \\
4:1 & 2,000 & 500 & 341,500 & 95,157 \\
1:1 & 1,250 & 1,250 & 214,919 & 228,823 \\
1:4 & 500  & 2,000 & 86,616 & 372,133 \\
1:9 & 250 & 2,250  & 43,482 & 419,853  \\ \bottomrule
\end{tabular}
\vspace{-1.em}
\label{table:training_data}
\end{table}

\noindent \textbf{Model Training.} All ResNet-34 encoders were trained on
the VoxCeleb2-GRC datasets with 40-dim log Mel filter bank features. Models were trained for 300 epochs on a single GPU with angular prototypical loss \cite{chung2020defence} and a minibatch of a 400 2-sec utterance segments. We used the Adam optimizer with the initial learning rate $0.001$ and a decay factor of 0.95 per epoch. The output embedding dimension is 512. 
The group-adapted encoders were obtained by fine-tuning the base encoder on the single-gender training subsets with the same training parameters for an additional 300 epochs. 
The score fusion model is a three-layer MLP, where each hidden layer has 32 units with ReLU activation. The fusion network outputs a single fused score based on a sigmoid function.
We trained the score fusion model with 200,000 randomly sampled positive and negative pairs constructed from VoxCeleb2. Positive and negative training pairs were combined and shuffled on-the-fly during the training, with a minibatch batch of 1000 3-sec segments, 50 training epochs, and 0.001 learning rate.
\\
\noindent \textbf{Evaluation.} False accept rate (FAR) is the fraction of imposter speakers' utterances that are falsely accepted; False rejection rate (FRR) is the fraction of true speakers' utterances that are falsely rejected. Equal error rate (EER) is the rate where FAR and FRR are equal. 
We mainly use EER to characterize the performance of speaker verification models on different subsets of speakers.  To probe the model fairness according to gender, we define \textbf{group-wise EERs}: $EER[F]$, where the EER is derived from FAR and FRR when the true speakers are female speakers; $EER[M]$ is defined correspondingly. We denote $EER[\text{All}]$ as the \textbf{overall EER}. Given the group-wise EERs, we can characterize the model unfairness across groups via the \textbf{disparity score (DS)} $DS = |EER[F] - EER[M]|.$ The usage of different trials for computing these EER metrics is depicted in Table~\ref{table:eval_data}. Cosine similarity scores or their fused version is used to compute EERs. The cosine similarity between two utterances is evaluated according to the protocol in \cite{chung2018voxceleb2}. For each pair, ten 3-second temporal crops are sampled from each utterance and used to compute the mean similarity between all crops. EERs are reported with unit \%.

\vspace{-1.2em}

\section{Results}

\vspace{-.5em}
\subsection{Model Fairness with Imbalanced Group Representation}
\vspace{-.5em}

We first investigate the impact of imbalanced training dataset on model fairness in typical deep speaker verification models. We consider two state-of-the-art deep neural networks~\cite{chung2020defence}, a quarter-channel ResNet-34 baseline (Q/RN) with 1.4M parameters and a larger half-channel ResNet-34 baseline (H/RN) with 5.6M parameters. The two baselines are trained on the VoxCeleb2-GRC datasets and evaluated on the VoxCeleb1-F dataset.

\begin{figure}[!htb]
    \centering
    \includegraphics[width=1.0\columnwidth]{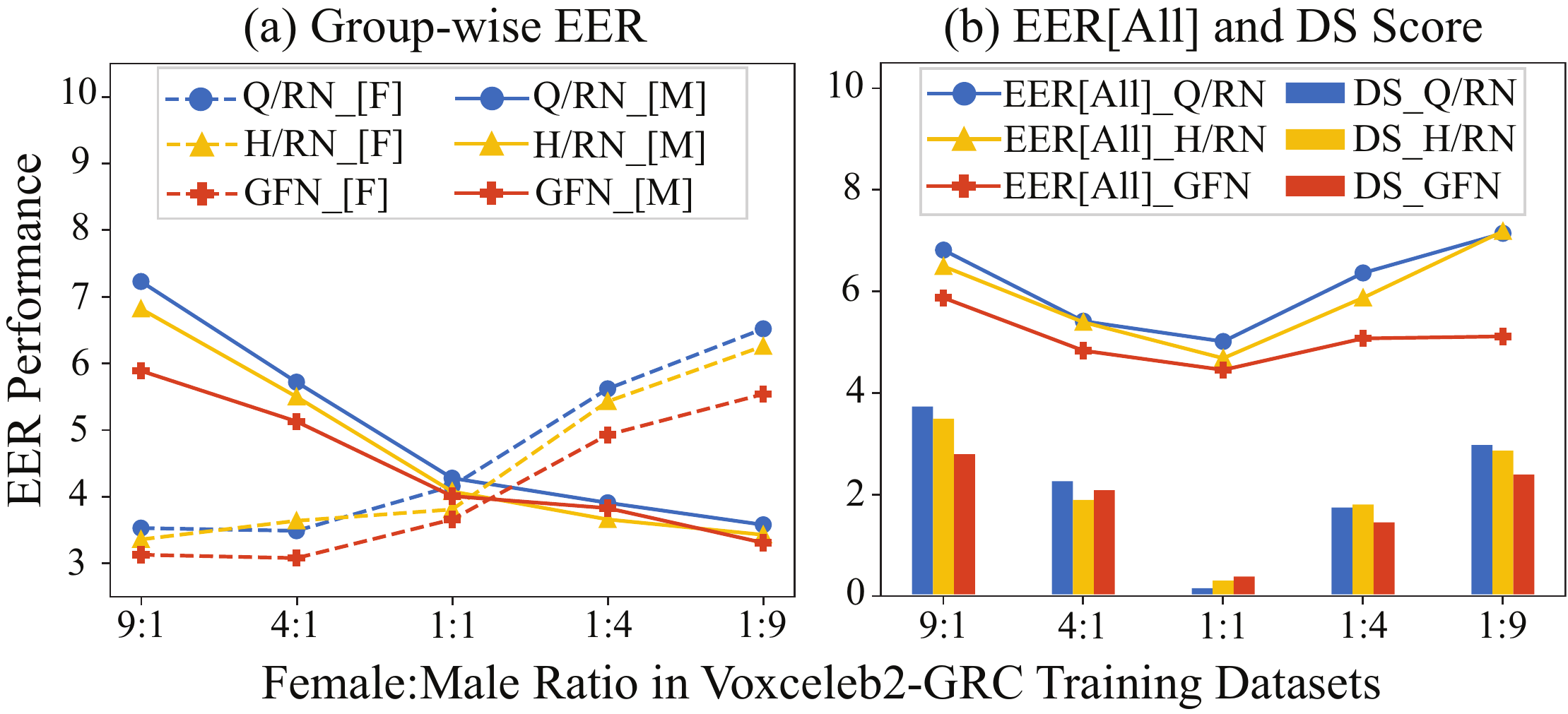}
    \caption{VoxCeleb1-F evaluation results from models trained on VoxCeleb2-GRC datasets.
    Q/RN, H/RN: baseline ResNet models; GFN: gender-adapted fusion of networks.
    }
    \label{fig:eval_results}
    \vspace{-1.8em}
\end{figure}

As showed in Figure~\ref{fig:eval_results},\footnote{\scriptsize All results in tabular form are also available on the website given in Footnote \ref{data_url}.} when the total number of speakers in training set are kept the same, the majority group has better group-wise EER than the minority group. 
This is, increasing dominance of one gender group ({\em e.g.,} from ratio 4:1 and 9:1) leads to increasing performance gap or model unfairness, indicated by the increasing $DS$ values ({\em e.g.,} from 1.71 to 3.70).
For example, when F:M = 9:1 in the training set, the Q/RN baseline has a female-group EER of 3.52, which is much better than the male-group EER of 7.22. A similar gap between female EER and male EER is observed in the setting of F:M = 1:9. When the training set is balanced, both baselines achieve roughly equal group-wise EER for both genders, indicated by the lowest $DS$ values ($<0.35$). Notably, overall EER increases as the training dataset becomes increasingly imbalanced, even though total training data sizes remain similar.\footnote{\scriptsize The differences between operating points for group-wise EERs and overall EER are less than 0.09.}

Among the two baselines, H/RN baseline achieves better group-wise and overall EERs among all gender ratios and smaller $DS$ than those of Q/RN. We attribute the fairness improvement to the larger learning capacity of the H/RN baseline. However, the H/RN baseline, which is four times of Q/RN baseline in model size, can only provide relatively small improvement on group-wise EERs and overall EER ($\approx$5\%).

\begin{table}[t!]
\caption{Trial statistics in VoxCeleb1-F evaluation datasets. 
}
\scriptsize
\centering
\begin{tabular}{rc|ccc}
\toprule
      \textbf{Gender Trials} & \textbf{Trial Count} & \multicolumn{3}{c}{\textbf{VoxCeleb1-F}} \\
        
        &  & [F] & [M] & [All] \\ \midrule
Positive F-F  & 150,000  & \checkmark &  & \checkmark \\
Negative F-F & 150,000   & \checkmark &  & \checkmark \\
Negative M-F & 150,000   & \checkmark & \checkmark & \checkmark \\
Positive M-M  & 150,000  & & \checkmark & \checkmark \\
Negative M-M & 150,000   & & \checkmark & \checkmark 
\\ \bottomrule
\end{tabular}
\label{table:eval_data}
\vspace{-1.2em}
\end{table}

\vspace{-.5em}
\subsection{Improving Fairness via Group-adapted Fusion Network}
\vspace{-.5em}

Now we consider the performance of the proposed group-adapted fusion network (GFN) model.  The GFN has 4.3M parameters, which is around 3 times that of Q/RN and smaller than H/RN.
Compared to the two baselines (in Figure~\ref{fig:eval_results}), the GFN achieves better group-wise and overall EERs regardless of gender group imbalance in the training sets.
In particular, GFN  achieves a female group EER of 3.12 (realizing 11.4\% and 6.9\% improvement relative to Q/RN and H/RN, resp.), and a male group EER of 5.88 (18.6\%/13.7\% improvement) in the F:M=9:1 setting.
For overall EERs, GFN achieves an EER of 5.84 (13.9\%/9.6\% improvement) and 5.08 (28.6\%/29.0\% improvement) in the F:M = 9:1 and 1:9 settings, respectively.

Note that GFN offers larger relative improvements for the minority group than for the majority group, thereby reducing the model unfairness or disparity score $DS$ in the imbalanced cases. Specifically, GFN achieves a $DS$ of 2.76 (25.4\%/20.2\% improvement relative to Q/RN and H/RN) and 2.23 (24.2\%/21.2\% relative improvement) in the F:M = 9:1 and 1:9 settings, respectively.

\vspace{-.5em}
\subsection{Embedding Visualization and Analysis}
\vspace{-.5em}

\begin{figure}[!t]
    \centering
    
    \includegraphics[width=1.0\columnwidth]{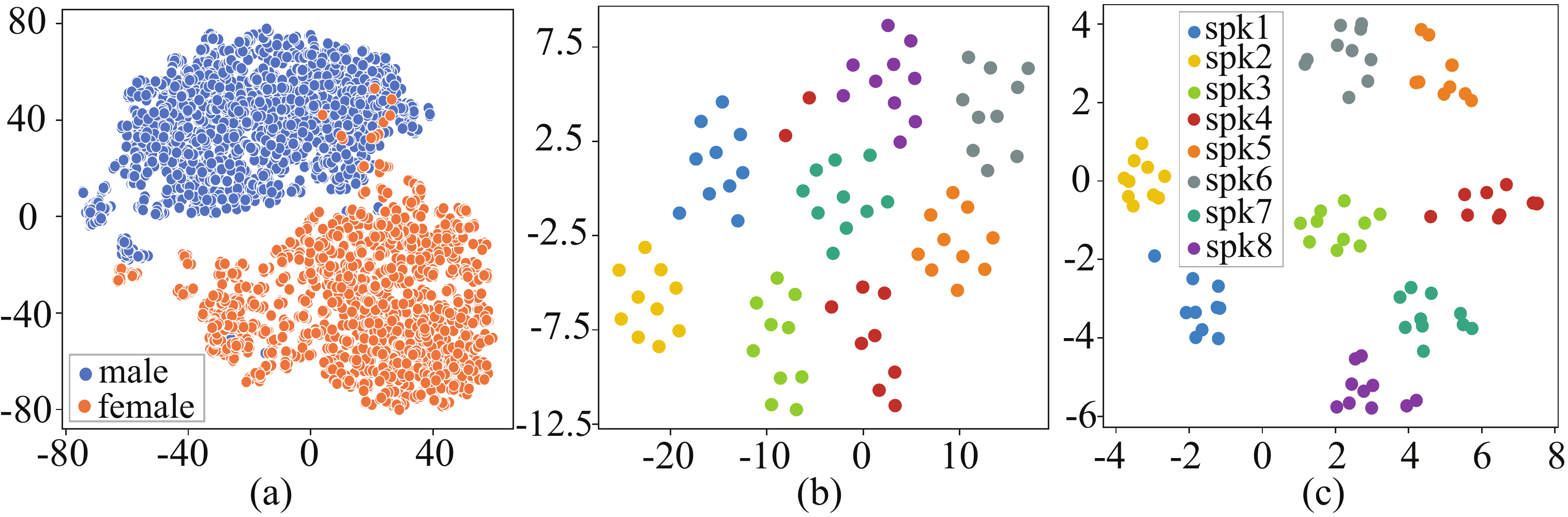}
    \caption{Visualization of learned speaker embeddings using the t-SNE method. (a) Male and female speaker utterance embeddings from Q/RN model; (b, c) Utterance embeddings of 8 randomly-sampled speakers from Q/RN model (b) and GFN model (c).
    }
    \label{fig:emb_vis}
    \vspace{-1.5em}
\end{figure}

To shed light on the cause of model unfairness and the effects of GFN modeling, we first use t-SNE to visualize utterance embeddings (10000 female and male base embeddings from Q/FN trained on F:M=1:1 dataset) in a 2D space. As shown in Figure~\ref{fig:emb_vis}(a), utterances of the same gender tend to aggregate in separate regions of the embedding space, which is consistent with the perception that same-gender voices sound relatively similar compared to between-gender differences. 
We hypothesize that in an imbalanced training setting, adapting encoders separately for different genders allows each encoder to improve the representation of different subregions of the embedding space, thus alleviating the bias toward the dominant group found in the single-model framework.

The t-SNE method can visualize the benefit of adapting and fusing group-specific embeddings by the clustering and separation of embeddings from different speakers. Figures~\ref{fig:emb_vis} (b) and (c) show the low-dimensional utterances embeddings from the baseline Q/RN encoder and from the concatenated three GFN embeddings, respectively. 
The adapted encoders tend to generate more compact speaker clusters with more separation between speakers, compared to the baseline encoder.
The compactness of speaker clusters can also be quantified via \textit{silhouette coefficients} (SC) \cite{rousseeuw1987silhouettes} used in clustering analysis, implemented by Scikit-learn \cite{scikit-learn}. SC measures how similar an utterance is to its own speaker cluster (compactness) compared to other speaker clusters (separation); a higher SC is better. We compute the SC from 80 utterance embeddings of 8 randomly-sampled speakers. The mean SC from the Q/RN baseline is 0.64, while the mean SC from the adapted encoder is 0.83, indicating that the adapted encoder extracts better embeddings for speaker differentiation purposes.

\vspace{-.5em}
\subsection{Ablation Studies}
\vspace{-.5em}

\begin{figure}[!t]
    \centering

    \includegraphics[width=0.70\columnwidth]{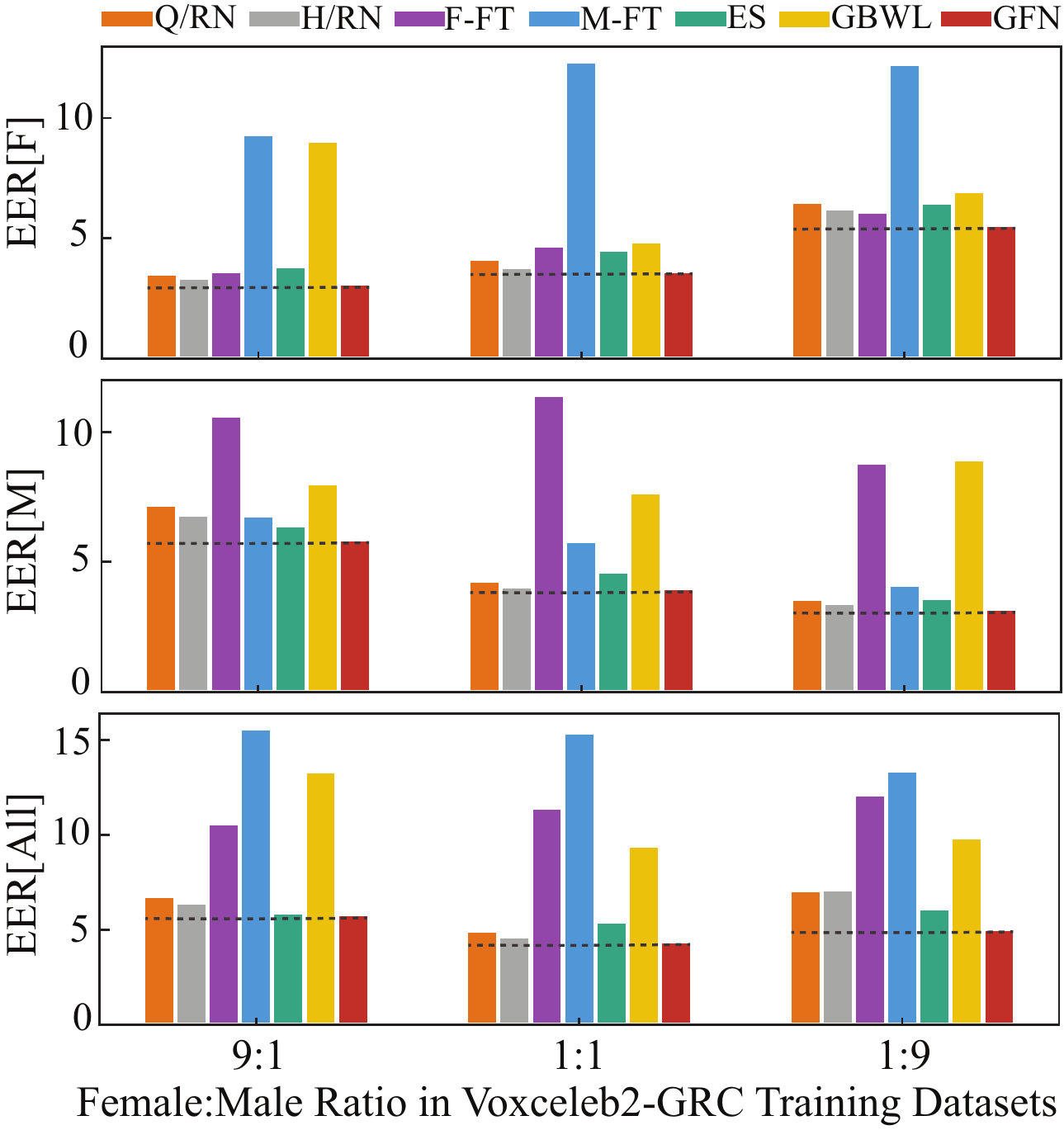}
    \caption{VoxCeleb1-F evaluation results from models trained on VoxCeleb2-GRC datasets under various ablation study conditions. Black dashed lines are results from our GFN models.
    }
    \label{fig:ablation_results}
    \vspace{-2.em}
\end{figure}

Several ablation studies reveal the benefits of different aspects of our model, as summarized in Figure~\ref{fig:ablation_results}. We first examine the impact of using only one adapted encoder without fusing other encoders.
Using a single male-adapted encoder (M-FT) typically degrades the overall EER, particularly the female group-wise EER although it might improve the male group-wise EER slightly when the male group is the minority. A similar phenomenon is observed when only using a female-adapted encoder (F-FT). Since male and female voices have distinct characteristics, we hypothesize that fine-tuning an encoder to one gender subset can cause the encoder to forget the learned features of the other gender. Therefore, it is necessary to fuse separately-adapted encoders to achieve better performance.  
To further verify the efficacy of the score fusion strategy we compare GFN's 3-layer MLP score fusion model with an equal-weight score (ES) fusion strategy (i.e., 1/3 for each input score). ES fusion achieves better group-wise EER than the baselines for gender-imbalanced settings, but gives overall worse performance than GFN.

Additionally, we consider an alternative embedding adaptation method named ``gender batching with weighted loss'' (GBWL), which fine-tunes the Q/RN base model by 1) alternating all-female and all-male mini-batches and 2) reciprocal weight for the majority-gender minibatch (\textit{i.e.}, when training on F:M=1:9, we scale the loss from all-male minibatches by 1/9). The GBWL method significantly degrades group-wise and overall EER in the baseline Q/RN (and similarly for H/RN). This shows the benefit and necessity of performing embedding adaptation using separate networks.

We also investigated the gated mixture-of-experts (MOE)~\cite{shazeer2017outrageously} strategy for fusing scores, which incorporates both adapted embeddings and similarity scores as inputs. However, gated-MOE achieves worse performance in the current score fusion training setting.

\vspace{-1.em}

\section{Conclusion}
We have analyzed the effect of imbalanced training data on group fairness of modern speaker verification models, by manipulating the gender balance in a VoxCeleb-based data set. The results show that there is a direct relationship between training set imbalance and verification accuracy on the test set, both overall and for the underrepresented group.
To improve performance fairness we developed a modular classifier architecture based on group-adapted encoders that are fused at the score level. 
Our approach achieves improvements in both overall and group-wise metrics, and reduces the performance gap between groups in scenarios with imbalanced training data. 
Specifically, our proposed method achieves relative reductions in overall EER of 9.6\% to 29.0\%, in minority group EER of 13.7\% to 18.6\%, and narrows EER disparity by 20.0\% to 25.4\%, compared to baselines.
Note that our approach can be generalized to more problematic scenarios, such as children and elderly demographic groups, by incorporating and fusing more group-adapted encoders. Additionally, robust backend scoring approaches such as PLDA  \cite{ferrer2022speaker,ramoji2020nplda} are worth exploring to alleviate model unfairness. Finally, performing group-adaptation and fusion on the training dataset might introduce additional overfitting risk, which should be taken into account when applying a trained GFN to out-of-domain datasets.

\vspace{-1.8em}
\section{Acknowledgments}
\vspace{-1em}
We thank our colleague Victor Rozgic and Alexa Speaker Understanding team members and managers for their input and support.

\bibliographystyle{IEEEbib}
\bibliography{refs}

\end{document}